\documentclass[letterpaper,aps,prl,superscriptaddress,twocolumn]{revtex4}

\usepackage{amsmath}
\usepackage{amssymb}
\usepackage{amsfonts}
\usepackage{relsize}
\usepackage{floatrow}

\usepackage{bm}
\usepackage{color}

\usepackage{graphicx}
\usepackage{epstopdf}

\newcommand{\thistext}{paper} 
\newcommand{\rd}{\ensuremath{\mathrm{d}}}
\newcommand{\sub}[1]{\ensuremath{_{\text{#1}}}}

\begin{document}
\title{Effect of bremsstrahlung radiation emission on fast electrons in plasmas \vspace{-2.2mm}}

\author{O. Embr\'eus}
\affiliation{Department of  Physics, Chalmers University of Technology, Gothenburg, SWEDEN}
\author{A. Stahl}
\affiliation{Department of Physics, Chalmers University of Technology, Gothenburg, SWEDEN}
\author{T. F\"{u}l\"{o}p}
\affiliation{Department of Physics, Chalmers University of Technology, Gothenburg, SWEDEN}
\date{\today}

\begin{abstract}  
\vspace{-3mm}
  Bremsstrahlung radiation emission is an important energy loss mechanism for
  energetic electrons in plasmas.  In this \thistext{} we investigate
  the effect of spontaneous bremsstrahlung emission on the
  momentum-space structure of the electron distribution, fully
  accounting for the emission of finite-energy photons. We find that
  electrons accelerated by electric fields can reach significantly higher energies than
  what is expected from energy-loss considerations. Furthermore, we
  show that the emission of soft photons can contribute significantly
  to the dynamics of electrons with an anisotropic distribution. 
\end{abstract}

\maketitle 
\noindent Energetic electrons are ubiquitous in
plasmas, and bremsstrahlung radiation is one of their most important
energy loss mechanisms \cite{betheheitler1934,blumenthal1970}.  At
sufficiently high electron energy, around a few hundred megaelectronvolts in
hydrogen plasmas, the energy loss associated with the emission of
bremsstrahlung radiation dominates the energy loss by collisions. 
Bremsstrahlung emission can also strongly
affect electrons at lower energies, particularly in plasmas containing
highly charged ion species.

An important electron acceleration process, producing energetic
electrons in both space and laboratory plasmas, is the runaway
mechanism \cite{dreicer1959}.
In the presence of an electric field which exceeds the minimum to overcome
 collisional 
friction~\cite{connorhastie}, a fraction of the charged particles can detach from the bulk
population and be accelerated to high energies, where radiative losses
become important.
Previous studies
{of laboratory plasmas~\cite{bakhtiariPRL2005,solisetal2007} and lightning discharges~\cite{gurevich2001}}
have shown that the energy carried away by  
bremsstrahlung radiation is important in limiting the energy of
runaway electrons. {The effect of bremsstrahlung
radiation loss on energetic-electron transport has also been considered in 
astrophysical plasmas, for example in the context of solar flares~\cite{miller1989}.}
However, only the average {bremsstrahlung} friction force on test particles has
been considered {in these studies}. 
In this \thistext, we present the first quantitative kinetic study of how
bremsstrahlung emission affects the runaway-electron distribution function.

Starting from the Boltzmann electron transport equation,  we derive a
collision operator representing bremsstrahlung radiation reaction, fully accounting for the
finite energy and emission angle of the emitted photons. We implement the operator
in a continuum kinetic-equation solver \cite{codepaper},
and use it to study the effect of bremsstrahlung on the distribution
of electrons in 2D momentum-space. We find significant differences in
the distribution function when 
bremsstrahlung losses are modeled with a Boltzmann equation (referred to 
as the ``Boltzmann'' or ``full'' bremsstrahlung model), compared to the model where only
the average friction force is accounted for (the ``mean-force'' model).
In the former model, the maximum energy reached by the
energetic electrons is significantly higher than is predicted by the latter. 
In previous treatments which considered average energy loss~\cite{bakhtiariPRL2005,gurevich2001,solisetal2007} or isotropic 
plasmas~\cite{blumenthal1970}, the emission of soft (low-energy) photons did not influence the electron motion. 
We show that in the general case, emission of soft photons contributes significantly to 
angular deflection of the electron trajectories.

 We will treat bremsstrahlung as
  a binary interaction (``collision'') between two charged particles,
 resulting in the emission of a photon~\cite{betheheitler1934}. 
We shall describe the effect of such collisions on the rate of change of the
distribution function $f_a(t,\,\ensuremath{\mathbf{x}},\,\ensuremath{\mathbf{p}})$ of some particle
species $a$ at time $t$, position \ensuremath{\mathbf{x}} and momentum \ensuremath{\mathbf{p}}, defined
such that $n_a(t,\,\ensuremath{\mathbf{x}}) = \int \rd \ensuremath{\mathbf{p}} \, f_a(t,\,\ensuremath{\mathbf{x}},\,\ensuremath{\mathbf{p}})$
is the number density of species $a$ at \ensuremath{\mathbf{x}}.
In what follows we suppress the time- and space
dependence of all functions, as the collisions will be assumed local in
space-time, and we shall consider only spatially homogeneous plasmas.

The collision operator $C^{\text{B}}_{ab}\{f_a,\,f_b\}$ describing the rate of
change of the distribution function due to bremsstrahlung interactions
between species $a$ and $b$ is given by $C^{\text{B}}_{ab} = (\partial
f_a/\partial t)_{c,ab} = \int (\rd n_a)_{c,ab}/\rd t \rd \ensuremath{\mathbf{p}}$, where
the differential change $(\rd n_a)_{c,ab}$ in the phase-space density due
to collisions in a time interval $\rd t$ is given by~\cite{relativistic_boltzmann, akama1970}
\begin{align}
\bigr(\rd n_a\bigr)_{c,ab} = f_a(\ensuremath{\mathbf{p}}_1) f_b(\ensuremath{\mathbf{p}}_2)
\bar{g}_{\text\o} \rd \bar{\sigma}_{ab} \rd\ensuremath{\mathbf{p}}_1 \rd \ensuremath{\mathbf{p}}_2 \rd t\nonumber\\
\hspace{.4cm} - f_a(\ensuremath{\mathbf{p}})f_b(\ensuremath{\mathbf{p}}')g_{\text\o} \rd
\sigma_{ab} \rd \ensuremath{\mathbf{p}} \rd \ensuremath{\mathbf{p}}' \rd t.
\label{eq:diff change}
\end{align}
\noindent Here, $\rd \sigma_{ab} =
\rd\sigma_{ab}(\ensuremath{\mathbf{p}}_1,\,\ensuremath{\mathbf{p}}_2,\,\ensuremath{\mathbf{k}};\,\ensuremath{\mathbf{p}},\,\ensuremath{\mathbf{p}}')$ is the
differential cross-section for a particle $a$ of momentum \ensuremath{\mathbf{p}} and a
particle $b$ of momentum $\ensuremath{\mathbf{p}}'$ to be taken to momentum $\ensuremath{\mathbf{p}}_1$
and $\ensuremath{\mathbf{p}}_2$, respectively, while emitting a photon of momentum
$\ensuremath{\mathbf{k}}/c$. We have also introduced the M\o{}ller relative speed
$g_{\text\o} = \sqrt{(\ensuremath{\mathbf{v}}-\ensuremath{\mathbf{v}}')^2 - (\ensuremath{\mathbf{v}}\times
  \ensuremath{\mathbf{v}}')^2/c^2}$. The barred quantities $\rd\bar\sigma$ and $\bar
g_{\text\o}$ are defined likewise, but with 
$(\ensuremath{\mathbf{p}},\,\ensuremath{\mathbf{p}}')$ and $(\ensuremath{\mathbf{p}}_1,\,\ensuremath{\mathbf{p}}_2)$ exchanged.  Eq.~(\ref{eq:diff change}) accounts only for the effect on the distribution of the
{spontaneous} emission of photons; interactions with existing photons by {absorption and stimulated bremsstrahlung emission will be neglected here. The correction to the collision operator by these processes is described in~\cite{oxenius1986}; the effect is negligible when $\phi(\ensuremath{\mathbf{x}},\,\ensuremath{\mathbf{p}}) \ll 2/h^3$, where $h$ is Planck's constant and $\phi$ is the distribution function of photons.
An estimate of the photon distribution function shows that the corrections are important for sufficiently dense, or large, plasmas; however, for the special case of electron runaway during tokamak disruptions, which is of particular concern, the corrections may be safely neglected. In other scenarios it is primarily bremsstrahlung processes involving low-energy photons that may be affected.}
The collision operator then takes the form
\begin{align}
C_{ab}^{\text{B}}(\ensuremath{\mathbf{p}}) &= \int \rd\ensuremath{\mathbf{p}}_1\, f_a(\ensuremath{\mathbf{p}}_1) \int \rd\ensuremath{\mathbf{p}}_2 \, \bar{g}_\text\o f_b(\ensuremath{\mathbf{p}}_2)\frac{\partial \bar\sigma_{ab}}{\partial \ensuremath{\mathbf{p}}} \label{eq:explicit collision operator}
\\
 &- f_a(\ensuremath{\mathbf{p}})\int\rd\ensuremath{\mathbf{p}}'\,g_\text\o f_b(\ensuremath{\mathbf{p}}')\, \sigma_{ab},
\nonumber
\end{align}
where $\sigma_{ab} = \int \rd \ensuremath{\mathbf{p}}_1 \, (\partial \sigma_{ab}/\partial
\ensuremath{\mathbf{p}}_1)$ is the total cross-section.  A significant simplification
to (\ref{eq:explicit collision operator}) occurs if
(i) target particles can be assumed
stationary, $f_b(\ensuremath{\mathbf{p}}) = n_b\delta(\ensuremath{\mathbf{p}})$; and (ii) the plasma is
cylindrically symmetric (and spin unpolarized), $f_a(\ensuremath{\mathbf{p}}) = f_a(p,\,\cos\theta)$, where
$\cos\theta = p_\parallel / p$ and $p_\parallel$ is the Cartesian
component of $\ensuremath{\mathbf{p}}$ along the symmetry axis. 
Then the differential cross-section $\partial
\bar{\sigma}_{ab}/\partial \ensuremath{\mathbf{p}}$, for
an electron to scatter from momentum \ensuremath{\mathbf{p}} into $\ensuremath{\mathbf{p}}_1$ with the 
emission of a photon, depends only on $p,\,p_1$ and $\cos\theta_s=
\ensuremath{\mathbf{p}}_1\cdot\ensuremath{\mathbf{p}}/p_1p$. 
The resulting operator can be conveniently
expressed in terms of an expansion in Legendre polynomials $P_L$. 
We write $f_a(\ensuremath{\mathbf{p}})= \sum_L
f_L(p)P_L(\cos\theta)$ and $C^{\text{B}}_{ab}(\ensuremath{\mathbf{p}}) = \sum_L C^{\text{B}}_L(p)
P_L(\cos\theta)$, and obtain
\begin{align}
\hspace{-1.8mm}C^{\text{B}}_L(p) &= n_b\hspace{-1mm}\int \hspace{-1mm}\rd p_1 \biggr[ p_1^2v_1f_L(p_1)\,2\pi \hspace{-1mm}\int_{-1}^1 \hspace{-1mm} \rd \cos\theta_s \nonumber \\
&\hspace{15mm} \times \hspace{-.8mm} P_L(\cos\theta_s)\frac{\partial \bar\sigma_{ab}}{\partial \ensuremath{\mathbf{p}}}\biggr] \hspace{-.9mm}- \hspace{-.5mm}n_b v f_L(p) \sigma_{ab}(p).
\label{eq:legendre boltz}
\end{align}
The integration limits in $p_1$ are determined by the conservation of
energy, giving $m_e c \sqrt{(\gamma+k/m_ec^2)^2-1} < p_1 < \infty$. In
this work we use the differential cross-section $\partial\bar\sigma/\partial
\ensuremath{\mathbf{p}}$ for scattering in a static Coulomb field in the Born
approximation, integrated over photon emission angles. This expression was first derived by Racah~\cite{racah1934}, with a misprint
later corrected in~\cite{McCormick1956}. For the Boltzmann model
this full cross-section is employed, while for the mean-force model we use
the high-energy limit as in~\cite{bakhtiariPRL2005,gurevich2001,solisetal2007}.

A useful approximation to the collision operator is obtained by
neglecting the deflection of the electron in the bremsstrahlung
reactions, formally achieved by the replacement $\partial
  \bar\sigma/\partial \ensuremath{\mathbf{p}} = [\delta(\cos\theta_s-1)/2\pi
  p^2]\partial \bar\sigma/\partial p$, where $\partial \bar\sigma/\partial p =
p^2\int \rd\Omega_s \partial \bar\sigma/\partial \ensuremath{\mathbf{p}} $, which yields
\begin{align}
C^{\text{B}}_{ab}(\ensuremath{\mathbf{p}}) &\approx n_b\int \rd p_1 \, v_1
f_a(p_1,\,\cos\theta)\frac{\partial \bar\sigma}{\partial p}(p;\,p_1) \nonumber \\
&- n_b v
f_a(p,\,\cos\theta)\sigma(p). 
\label{eq:delta cos}
\end{align}
This takes the form of a one-dimensional integral operator acting only on the energy
variable, and involves the  integrated cross-section 
which is well known and given analytically for example in Eq.~(14) of
\cite{betheheitler1934}. Physically, this approximation is
motivated by the strong forward-peaking of the cross-section, with
typical deflection angles being of order $\theta_s \sim 1/\gamma$ due to
relativistic beaming.

The brems\-strahlung cross-section has an infrared divergence; for low
photon energies $k$, it diverges logarithmically as $\rd \sigma
\propto 1/k$. The total energy loss rate is however finite, indicating
that a large number of photons carrying negligible net energy are
emitted. A consequence of this behavior is that the two terms in the
Boltzmann operator (\ref{eq:explicit collision operator}) are individually infinitely large, necessitating
the introduction of a photon cut-off energy $k_0$, below which the
bremsstrahlung interactions are ignored in~(\ref{eq:legendre boltz}) and (\ref{eq:delta cos}).
We can however proceed analytically to evaluate the effect of the
low-energy photons.  While they carry little energy, they may
contribute to angular deflection, analogously to the small-angle
collisions associated with elastic scattering. Taylor expanding
(\ref{eq:legendre boltz}) in small photon energy $k =
\gamma_1-\gamma$ yields to leading order
\begin{align}
C_L^{\text{\,small-$k$}} = -n_b v f_L(p) \int_{k_c}^{k_0} \rd k \, \int_{-1}^1 \rd\cos\theta_s \, \nonumber\\\times\Big[1-P_L(\cos\theta_s)\Big]  \frac{\partial \bar\sigma}{\partial k \partial\cos\theta_s}.
\label{eq:small k boltz}
\end{align}
Since $P_0(\cos\theta_s) \equiv 1$, the angle-averaged electron distribution
(represented by the $L=0$ term) is not 
directly affected by the low-energy photons, reflecting the fact that
the photons carry negligible energy, consistent with the description by Blumenthal \& Gould~\cite{blumenthal1970}
for the isotropic case. Due to the logarithmic divergence
of the cross-section, however, a significant contribution to
angular deflection (represented by the $L\neq 0$ terms) is possible.  Inspection of the integrand in 
(\ref{eq:small k boltz}) further reveals that significant contributions
originate from large-angle scatterings, indicating that a Fokker-Planck
approximation is inappropriate. While it may seem counter-intuitive that 
low-energy photon emissions contribute to large-angle collisions, note
that due to the large mass ratio between electron and ion, large
momentum transfers to the nucleus is allowed even without energy transfer.
When the electron energy exceeds the ion rest energy, however, ion recoil
effects will modify (\ref{eq:small k boltz}).
 
We can quantify the importance of the low-energy photons by
calculating the $L=1$ term of (\ref{eq:small k boltz}) --
giving the loss rate of parallel momentum -- and comparing it to the
corresponding term of the elastic-scattering collision operator given in~\cite{codepaper}. 
Carrying out the integration, one obtains the ratio
\vspace{-.2mm}
\begin{align}
\hspace{-1mm}\frac{C_1^{\text{small-$k$}}}{C_1^{\text{elastic}}} = \alpha \frac{2}{\pi}\frac{\ln\Lambda\sub{B}}{\ln\Lambda} \biggr[\hspace{-1mm}\left(\ln\frac{2p}{m_ec}\right)^2 \hspace{-1.5mm} - 2\ln\frac{2p}{m_ec}+ 2\biggr],
\label{eq:ratio}
\end{align}
with a relative
error of magnitude $O( m_e^2c^2/p^2) + O(k_0/pc)$, and where 
$\alpha = e^2/4\pi\varepsilon_0 \hbar c \simeq 1/137$ is the 
fine-structure constant. We have introduced a
bremsstrahlung logarithm $\ln \Lambda\sub{B} = \ln(k_0/k_c)$, which
arises in a way similar to the Coulomb logarithm $\ln\Lambda$ for elastic
collisions, and is due to cutting off the logarithmic divergence at
some lowest photon energy $k_c$.  This energy corresponds to photons
emitted at the plasma frequency $\omega_p$, at which point polarization of the
background will dampen the bremsstrahlung
interactions~\cite{galitsky1963}, and is thus given by $k_c =
\hbar\omega_p$. This gives a bremsstrahlung logarithm $\ln \Lambda\sub{B}
\approx 21 + \ln\left(k_0/(m_ec^2\sqrt{n_{20}})\right)$, where $n_{20} = n_e/(10^{20}\,\text{m}^{-3})$ is the electron density in units of $10^{20}\,$m$^{-3}$.
Assuming a plasma with $\ln\Lambda = 15$, $n_{20} = 1$
and choosing $k_0 = 0.01p$, the ratio~(\ref{eq:ratio}) is of order 10\% at 30\,MeV,
50\% at 2\,GeV and 100\% at 30 GeV, demonstrating that {angular}
deflection caused by the emission of low-energy photons can contribute
significantly to the motion of highly energetic electrons.

The
bremsstrahlung collision operator has been implemented in the
initial-value continuum kinetic-equation solver \textsc{CODE}
(COllisional Distribution of Electrons)~\cite{codepaper}. For this
study we use \textsc{CODE} to solve the  equation
\begin{align}
\frac{\partial f_e}{\partial t} - e E_\parallel \frac{\partial f_e}{\partial p_\parallel} = C^{\text{FP}}\{f_e\} + C^{\text{B}}\{f_e\}, 
\label{eq:kinetic eq}
\end{align}
which in a magnetized plasma represents the gyro-averaged kinetic equation,
with the parallel direction given by the magnetic field \ensuremath{\mathbf{B}}.
The equation is also valid for an unmagnetized plasma which is cylindrically symmetric
around the electric field \ensuremath{\mathbf{E}}.
Elastic collisions are accounted for by the linearized
relativistic Fokker-Planck operator for Coulomb collisions
$C^{\text{FP}}$, and $C^{\text{B}}$ is the bremsstrahlung operator
$C^{\text{B}}_{eb}$ summed over all particle species $b$ in the plasma.
Both thermal and fast electrons are resolved
simultaneously, allowing runaway generation as well as the slowing-down 
of the fast population to be accurately modeled.

We will compare the effect of bremsstrahlung radiation losses on the
momentum-space distribution of fast electrons using several models. The
contribution from the emission of large-energy photons (with $k>k_0$)
are accounted for by either the Boltzmann operator in~(\ref{eq:legendre
  boltz}) or its approximation without angular deflection
(\ref{eq:delta cos}), while the low-energy photon contribution ($k<k_0$)
is described by~(\ref{eq:small k boltz}). For the numerical solutions we choose
an energy-dependent cut-off 
$k_0 = m_e c^2 (\gamma-1)/1000$. The Boltzmann models will be compared to
the mean-force
model where the bremsstrahlung losses are accounted for by an isotropic 
force term in the kinetic equation, defined as 
$\ensuremath{\mathbf{F}}\sub{B} = -\hat{\ensuremath{\mathbf{p}}}\sum_b n_b \int_0^{m_ec^2(\gamma-1)} \rd k
\, k \partial \sigma_{eb}/\partial k$, which is chosen to produce the correct
average energy-loss rate~\cite{betheheitler1934}.

\begin{figure}[htbp]
\begin{center}
\includegraphics[width=\textwidth,trim=7mm 60mm 16mm 6mm]{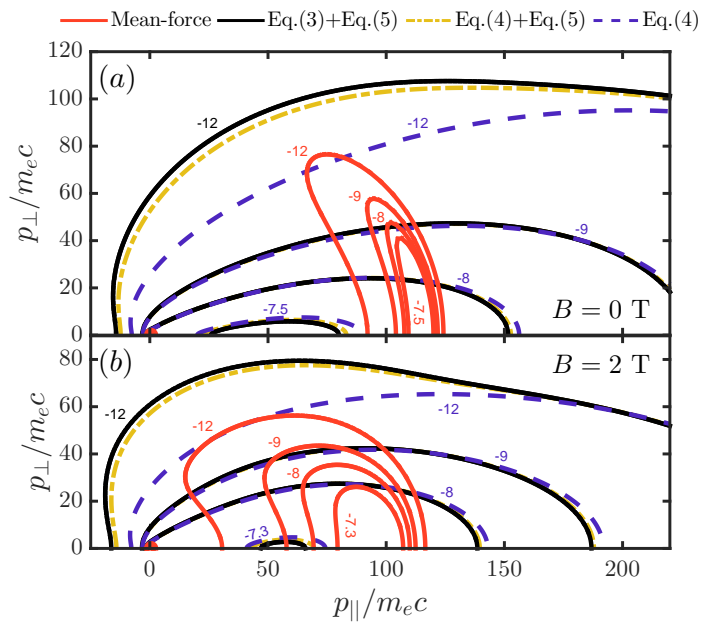}\vspace{-2mm}
 \caption{\label{fig:2D dist}{Steady-state electron 2D distribution functions; $(a)$ with no magnetic field, $(b)$ with {$B=2$\,T}. Electric field {$E  = 2E_c$}, plasma parameters {$Z\sub{eff} = 10$}, {$n_{20} = 30$} and $T_e = 10\,$keV. Contours show $\log_{10} F$, where $F = (2\pi m_e T_e)^{3/2} f_e / n_e$.}}\end{center}
\end{figure}

To characterize the effect of bremsstrahlung on the electron
distribution, we investigate quasi-steady-state numerical solutions of the
kinetic equation~(\ref{eq:kinetic eq}). 
These are obtained by evolving the distribution function in time until an equilibrium is reached, 
typically after a few seconds at density $n_{20} = 1$ if 
an initial seed of fast electrons is provided.
We investigate a range
of electric-field values near the minimum field $E_c = 4\pi \ln\Lambda n_e r_0^2 m_e c^2/e$
to overcome collisional friction~\cite{connorhastie}, {using plasma parameters characteristic
of tokamak-disruption experiments with massive gas injection}.

Figure \ref{fig:2D dist} shows the electron distribution function in
momentum space, calculated using \textsc{CODE}, with full Boltzmann
bremsstrahlung effects included (black, solid);
neglecting angular deflections in the large-$k$ contribution (yellow, dash-dotted);
also neglecting the small-$k$ contribution (blue, dashed); and finally using the
mean-force model (red, solid). 
Non-monotonic features (bumps) form in the mean-force as well as the 
Boltzmann models, but their characteristics are significantly different. 
With the Boltzmann models, an extended tail forms in the electron
distribution. In contrast, the mean-force model produces a sharp feature,
located where the energy gain due to the electric-field acceleration
balances friction and bremsstrahlung losses. The addition of low-$k$ 
scatterings (\ref{eq:small k boltz}), which lead to large-angle deflections, causes a subpopulation of fast
electrons with significant perpendicular momentum to form. Furthermore, (\ref{eq:legendre boltz})
and (\ref{eq:delta cos}) appear to generally produce the same qualitative features,
indicating that scatterings involving large-energy photons are well approximated by 
neglecting the angular deflection of the electron.

Inclusion of synchrotron radiation losses associated with the
gyromotion of electrons in a straight magnetic field has been shown to be an important energy-loss 
mechanism~\cite{andersson2001,stahl2015,hirvijoki2015,decker2016,pavel2015}.
Figure \ref{fig:2D dist}(b) shows that, in conjunction with bremsstrahlung losses, synchrotron losses (modelled as in
\cite{hirvijoki2015}) shifts the distribution
towards lower energies but does not change its qualitative features.
 The difference between the Boltzmann and mean-force models is
therefore reduced in such cases, as the extent of the distribution when
full bremsstrahlung effects are included is reduced by the synchrotron effect.

\begin{figure}[t]
\begin{center}
\includegraphics[width=\textwidth,trim=7mm 6mm 10mm 5mm]{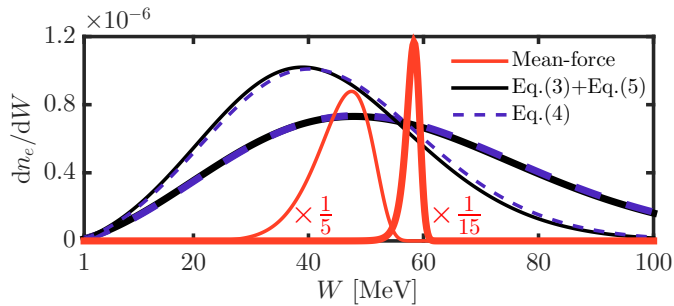}\vspace{-2mm}
 \caption{\label{fig:1D dist} Angle-averaged tail of the electron distributions in figure~\ref{fig:2D dist}, with $B=0$ (thick lines) and {$B =2\,$T} (thin lines). \vspace{-5mm}}
\end{center}
\end{figure}

Angle-averages of the electron distribution functions in figure~\ref{fig:2D dist} are 
shown in figure~\ref{fig:1D dist} as a function of electron kinetic energy $W = m_ec^2(\gamma-1)$. 
When there are no synchrotron losses present, the
difference between the Boltzmann models for bremsstrahlung losses is
insignificant. In the presence of effects which are sensitive to the 
angular distribution of electrons, such as 
synchrotron radiation losses (which are proportional to $p_\perp^2$), the
difference is somewhat enhanced as angular deflection
amplifies the dissipation.

\begin{figure}[t]
\begin{center}
\includegraphics[width=\textwidth,trim=4mm 3mm 6mm 4.5mm]{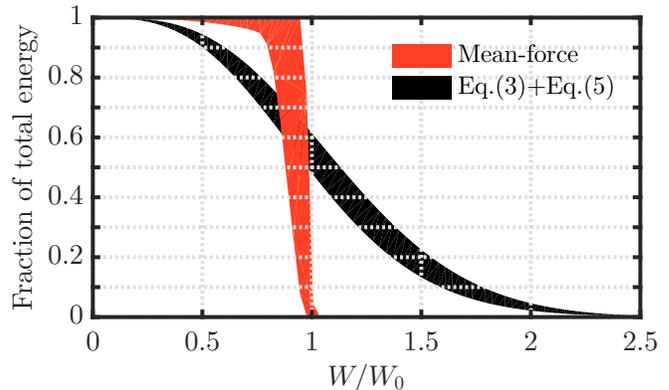}
\vspace{-8mm} \caption{\label{fig:Energy comparison2} {{Fraction of plasma kinetic energy carried by electrons of energy greater than $W$. The values are calculated from numerical solutions of the kinetic equation, and are shown as a function of normalized electron kinetic energy. Filled regions indicate the values spanned when $Z\sub{eff}$ and the normalized electric field $(E/E_c-1)/(Z\sub{eff}+1)$ are varied between 1 and 35, and 0.05 and 0.25 respectively.}\vspace{-3mm}}}
\end{center}
\end{figure}

{
To quantify the width in energy of the fast-electron tail, figure~\ref{fig:Energy comparison2} shows the fraction of total plasma kinetic energy carried by electrons with energy greater than $W$, for a range of plasma compositions and electric fields. Again, the steady-state solutions are considered, and the energy ratio is calculated as $\int_W^\infty \rd W \, W(\rd n_e/\rd W) / W\sub{tot}$. When normalized to the energy $W_0$ which solves the energy-balance equation $eE_\parallel - eE_c + F\sub{B}= 0$ (accounting for collisional and bremsstrahlung energy loss), the behavior is seen to be insensitive to electric field and effective charge. The Boltzmann model consistently predicts that a fraction of the electron population reaches significantly higher energies than in the mean-force model, where all electrons have energy near $W_0$. For instance, in the Boltzmann model 5\% of the plasma energy is carried by electrons with energy more than $2W_0$. }

\noindent{\em Summary}\ ---\ We have developed a kinetic description of the effect of spontaneous
bremsstrahlung emission on energetic electrons in plasmas. A
computationally efficient representation of the bremsstrahlung
collision operator has been obtained using an expansion in Legendre
polynomials, with which the operator is reduced to a set of 1D energy
integrals. This allows for rapid evaluation of the self-consistent
electron distribution function in the presence of bremsstrahlung
losses derived from the full Boltzmann operator.

By treating bremsstrahlung emission as a discrete process, we have
shown that electrons may be accelerated to significantly higher
energies than would be predicted by energy balance alone, with a
significant fraction of particles reaching {at least} twice the expected energy. {
The explanation for this can be intuitively understood in the single-particle picture, where the 
new model allows some electrons to suddenly lose a large fraction of  their energy in 
one emission, whereas others may be accelerated for a long time before a 
bremsstrahlung reaction occurs, thereby allowing higher maximum energies to be reached.}
This has important implications for the interpretation of experimental
observation of fast electron beams in plasmas where bremsstrahlung
losses are important, {such as in magnetic-fusion plasmas}.  
Furthermore, new effects are revealed in our treatment, as
the emission of soft photons is found to contribute to angular
deflection of the electron trajectory at a rate that increases with
electron energy. This effect shifts part of the momentum-space
distribution function towards higher perpendicular momenta, which in
turn has implications for e.g.~the destabilization of kinetic
instabilities or the level of synchrotron radiation loss in
magnetized plasmas.


\begin{acknowledgments}
The authors are grateful to  G.~Papp for fruitful discussions. This project has received funding from the European Union's Horizon 2020 research and innovation programme under grant agreement number 633053. The views and opinions expressed herein do not necessarily reflect those of the European Commission. This work was supported by the Knut and Alice Wallenberg Foundation (\textsc{PLIONA} project) and the European Research Council (ERC-2014-CoG grant 647121). 
\end{acknowledgments}

\bibliographystyle{unsrt}

\end{document}